\documentclass[preprint,12pt]{elsarticle}
\usepackage{amsmath,amsthm}
\newtheorem{theorem}{Theorem}
\newtheorem{proposition}{Proposition}

\usepackage{lineno,hyperref}
\usepackage{graphicx}
\usepackage{bm}
\usepackage{mathrsfs}
\usepackage{amsfonts}

\usepackage{epsfig}
\usepackage{caption}
\usepackage{bm}
\usepackage{mathrsfs}
\usepackage{pifont}
\usepackage{makecell}
\usepackage{natbib}
\usepackage{setspace}
\usepackage{geometry}
\usepackage{array,color}
\usepackage{appendix}
\usepackage{amssymb}
\usepackage{latexsym}
\usepackage{lineno,hyperref}
\usepackage{graphicx}
\usepackage{bm}
\usepackage{mathrsfs}
\usepackage{amsmath}
\usepackage{amsfonts}
\usepackage{epsfig}
\usepackage{caption}
\usepackage{floatrow}
\usepackage[percent]{overpic}
\usepackage{adjustbox}
\usepackage{bm}
\usepackage{mathrsfs}
\usepackage{pifont}
\usepackage{makecell}
\usepackage{natbib}
\usepackage{setspace}
\usepackage{geometry}
\usepackage{array,color}
\usepackage{appendix}
\usepackage{amssymb}
\usepackage{latexsym}
\usepackage{graphicx}
\usepackage{subfig}
\usepackage{float}
\usepackage{bm}
\usepackage{epstopdf}
\usepackage{multirow}
\usepackage{booktabs}
\usepackage{array}
\usepackage{hyperref}

\usepackage{mathrsfs}
\usepackage{amsfonts}
\usepackage{epsfig}
\usepackage{caption}
\usepackage{bm}
\usepackage{mathrsfs}
\usepackage{pifont}
\usepackage{makecell}
\usepackage{natbib}
\usepackage{setspace}
\usepackage{geometry}
\usepackage{array,color}
\usepackage{appendix}
\usepackage{epstopdf}

\modulolinenumbers[5]

\journal{Chaos, Solitons $\&$ Fractals}

\graphicspath{{pic/}}

\begin{document}

\begin{frontmatter}
	
	\title{Coevolutionary dynamics of feedback-evolving games in structured populations}

	\author[first]{Qiushuang Wang}
	\author[first]{Xiaojie Chen\corref{mycorrespondingauthor}}
	\cortext[mycorrespondingauthor]{Corresponding author}
	\ead{xiaojiechen@uestc.edu.cn}
	\author[third]{Attila Szolnoki}
	
	\address[first]{School of Mathematical Sciences, University of Electronic Science and Technology of China, Chengdu 611731, China}
	\address[third]{Institute of Technical Physics and Materials Science, Centre for Energy Research, P.O. Box 49, H-1525 Budapest, Hungary}

\begin{abstract}
The interdependence between an individual strategy decision and the resulting change of environmental state is often a subtle process. Feedback-evolving games have been a prevalent framework for studying such feedback in well-mixed populations, yielding important insights into the coevolutionary dynamics. However, since real populations are usually structured, it is essential to explore how population structure affects such coevolutionary dynamics. Our work proposes a coevolution model of strategies and environmental state in a structured population depicted by a regular graph. We investigate the system dynamics, and theoretically demonstrate that there exist different evolutionary outcomes including oscillation, bistability, the coexistence of oscillation and dominance, as well as the coexistence of cooperation and defection. Our theoretical predictions are validated through numerical calculations. By using Monte Carlo simulations we examine how the number of neighbors influences the coevolutionary dynamics, particularly the size of the attractive domain of the replete environmental state in the cases of bistability or cooperation-defection coexistence. Specifically, in the case of bistability, a larger neighborhood size may be beneficial to save the environment when the environmental enhancement rate by cooperation / degradation rate by defection is high. Conversely, if this ratio is low, a smaller neighborhood size is more beneficial. In the case of cooperator-defector coexistence, environmental maintenance is basically influenced by individual payoffs. When the ratio of temptation minus reward versus punishment minus sucker's payoff is high, a larger neighborhood size is more favorable. In contrast, when the mentioned ratio is low, a smaller neighborhood size is more advantageous.
\end{abstract}
	

{\tiny }\end{frontmatter}

\section{Introduction}

The environment plays a crucial role in supporting the survival and reproduction of humans. For example, the earth serves as the home of humanity, providing essential resources such as air, water, food, and other necessities~\cite{Vitousek_Science_1997,Mekonnen_SA_2016}. Consequently, it is essential for individuals to protect the environment, enhance energy efficiency, and promote sustainable development. Cooperative behavior  is essential for fostering the development of both human societies and the environment \cite{Liu_Ambio_2007,Karl_Science_2003,Perc_PR_2017,Pacheco_LR_2014,Rand_TCS_2013,Tu_NS_2023}. Understanding the emergence and maintenance of cooperation is a major  challenge, and evolutionary game theory provides a powerful framework to address this issue. \cite{Stewart_PNAS_2014,Axelrod_Science_1981,Taylor_MB_1978,Nowak_Nature_2004,Liu_Chaos_2018}.

In evolutionary game theory, replicator dynamics is the most commonly utilized approach to describe the evolutionary process of cooperation in an infinite well-mixed population \cite{Schuster_JTB_1983,Hofbauer_BAMS_2003,Wang_AMC_2018,Wang_PLA_2021,Hofbauer_Cambridge_1998,Nowak_Science_2004}.
Nevertheless, populations frequently exhibit a structural characteristic where individuals only interact with their immediate neighbors, reflecting a common pattern observed in various social and biological systems \cite{Bhalla_Science_1999,Koch_Science_1999,Wasserman_Cambridge_1994}.
Consequently, investigating the evolutionary dynamics of cooperative behavior in structured populations has emerged as a key research topic \cite{Abramson_PRE_2001,Szabo_PR_2007,Roca_PRE_2009,Ohtsuki_Nature_2006,Allen_Nature_2017,Xia_PLE_2023,Zhang_IEEETNSE_2025,Capraro_NCS_2024}.
For instance, the replicator equation on regular graphs has been developed to describe how the frequencies of different strategies evolve in an infinite structured population \cite{Ohtsuki_JTB_2006,Wang_NC_24}.

In practice, the evolution of strategies is not only determined by interactions among individuals within a population, but is also closely related to the environmental state \cite{Menge_PNAS_2008,Gore_Nature_2009,Innes_SP_2013,szolnoki_epl17,Estrela_TEE_2019}. This interplay indicates that the environment and strategies influence each other reciprocally, shaping the evolutionary dynamics.
Taking the example of humans and the natural world, the environment provides the essential foundation for human survival, influencing human behavior. Simultaneously, human activities impact the environment, creating a feedback loop between strategic behaviors and the environmental state.
Henceforth, comprehending and analyzing the interdependent coevolution of strategic actions and the environmental state is essential for studying complex systems \cite{Ebel_PRE_2002,Traulsen_PRL_2005,Traulsen_PRE_2006}.
Feedback-evolving game  provides an important framework for characterizing this coevolution \cite{Obradovich_JRSI_2019,Chen_PLB_2018,Yan_NJP_2021,Wang_EL_2020,Tilman_NC_2020}. Moreover, there have been many meaningful studies about the coevolutionary dynamics in well-mixed populations \cite{Hennes_AAMAS_2009,Weitz_PNAS_2016,Gong_Auto_2022,Pal_NC_2022,Levy_SNCS_2023,Arefin_PRSA_21,Jiang_PLOSCB_23,Hua_PRR_24,Liu_Elife_2023,Hua_ESWA_2023,Hua_ChaosSF_2023}. It is also significant to explore how the structural characteristics of the population affect the coevolutionary dynamics. However, until now few studies have addressed this issue.

In this paper, based on the replicator equation in structured populations \cite{Ohtsuki_JTB_2006}, we establish a coevolutionary model for feedback-evolving games in a structured population where each individual has an equal number of neighbors.
Through theoretical analysis and numerical simulations, we examine the coevolutionary dynamics of cooperation and the environmental state, revealing that these dynamics are significantly more complicated in structured populations compared to well-mixed populations.
Subsequently, we focus on cases where the system exhibits either bistability or the coexistence of cooperation and defection. In the situation of bistability, two stable equilibria emerge, corresponding to replete and depleted environmental states respectively.
Using Monte Carlo simulations, we investigate how the size of the basin of attraction for the replete environment equilibrium changes as the number of neighbors increases, highlighting the significant influence of the structural factor on the environmental maintenance.
We accordingly obtain that when the system exhibits bistability, a larger neighborhood size is beneficial for the environmental maintenance of the system if the ratio of the cooperators' enhancement rate to the defectors' degradation rate is high. Conversely, a smaller neighborhood size is more favorable if the ratio of the cooperators' enhancement rate to the defectors' degradation rate is low.
In the case of the coexistence of cooperation and defection, a larger neighborhood size is more favorable when the ratio of temptation minus reward to punishment minus sucker's payoff is close to zero (but still negative). However, a smaller neighborhood size is more beneficial when the ratio of temptation minus reward to punishment minus sucker's payoff is more negative.

\section{Model}\label{model}

We consider an  infinite structured population represented by a regular graph with degree $k$ ($ k>2 $ and $k\in\mathbb{N}$). Each individual corresponds to a vertex in the graph and has $k$ neighbors. It can choose to cooperate ($C$) or defect ($D$)  in a two-player feedback-evolving game with their neighbors. The individual's strategy is updated according to the rule of birth-death updating \cite{Ohtsuki_Nature_2006}. Specifically, in each evolutionary update step, one individual is chosen for reproduction based on their fitness from the entire population, and their offspring replaces a randomly selected neighbor. Hence the evolutionary dynamics of strategy frequency can be captured by the replicator equation on a graph \cite{Ohtsuki_JTB_2006}. In this context, we investigate the coupled evolution of strategies and the environment in such a structured population. To do that, let $x$ be the frequency of cooperators in the population. The environmental state, denoted as $m\in[0,1]$, is time-dependent and modified by strategies through the function $f(x)=\theta x-(1-x)$. Here, $\theta>0$ signifies the ratio of environmental enhancement rates to degradation rates of cooperators and defectors, respectively \cite{Weitz_PNAS_2016}. Consequently, the dynamical equation of the environment is given by
$
\dot{m}=m(1-m)f(x).
$
Besides, we assume that the environment-dependent payoff matrix is represented as
\begin{equation}
	\begin{split}
		A(m)
		=\ &(a_{ij}(m))_{2\times 2}\\
		=\ &(1-m)\begin{bmatrix}
			T & P\\
			R & S
		\end{bmatrix}
		+m
		\begin{bmatrix}
			R & S\\
			T & P
		\end{bmatrix}\\
		=\ &
		\begin{bmatrix}
			(1-m)T+mR &(1-m)P+mS\\
			(1-m)R+mT & (1-m)S+mP
		\end{bmatrix},
	\end{split}
\end{equation}
where $R$ is the reward for mutual cooperation, $T$ is the temptation to defect, $P$ is the punishment for mutual defection, and $S$ is the sucker's payoff of a cooperator who interacts with a defector, under the condition $ m=1$.
Accordingly, the coevolutionary dynamics for feedback-evolving games in a structured population can be written as
\begin{equation}\nonumber
	\left\{
	\begin{aligned}
		\epsilon \dot{x}&=x(1-x)\left[(f_{1}(x,m)-f_{2}(x,m))+b_{12}(m) \right], \\
		\dot{m}&=m(1-m)[\theta x-(1-x)],
	\end{aligned}
	\right.
\end{equation}
where $f_{i}(x,m)=xa_{i1}(m)+(1-x)a_{i2}(m)$ is the fitness of strategy $i$ ($ i=1,2 $), $\epsilon$ denotes the relative speed by which individual actions modify the environmental state, and $b_{ij}(m)$, an environment-dependent term, is defined as
$$
b_{12}(m)=\frac{a_{11}(m)+a_{12}(m)-a_{21}(m)-a_{22}(m)}{k-2}.
$$
Consequently,
\begin{equation}\label{rep_eq}
	\left\{
	\begin{aligned}
		\dot{x}&=\frac{1}{\epsilon}x(1-x)(1-2m)\left[\delta_{PS}+(\delta_{TR}-\delta_{PS})x+\frac{\delta_{PS}+\delta_{TR}}{k-2} \right], \\
		\dot{m}&=m(1-m)[\theta x-(1-x)],
	\end{aligned}
	\right.
\end{equation}
where $\delta_{PS}=P-S$ and $\delta_{TR}=T-R$.
As $ k\to\infty$, the system equation in structured populations, as described by  Eq. \eqref{rep_eq}, reverts to that in well-mixed populations \cite{Weitz_PNAS_2016}.
In the following sections, we will explore the coevolutionary dynamics of cooperation and environmental state in structured populations by means of theoretical analysis and numerical calculations.

\section{Theoretical Results}\label{result}

After calculations, it is determined that when  $1-k<\frac{T-R}{P-S}<\frac{1}{1-k}$, the system \eqref{rep_eq} has seven equilibria. These include four corner equilibria: $(0,0)$, $(0,1)$, $(1,0)$, and $(1,1)$; two boundary equilibria: $\left(x^{*},0\right)$ and $\left( x^{*},1\right) $, where $x^{*}\triangleq \frac{(1-k)(P-S)-(T-R)}{(k-2)(T-R-P+S)} $; and one interior equilibrium: $\left(\frac{1}{1+\theta},\frac{1}{2}\right)$.
When $T-R>{\rm max}\left\lbrace\frac{P-S}{1-k},(1-k)(P-S) \right\rbrace $ or $T-R<{\rm min}\left\lbrace\frac{P-S}{1-k},(1-k)(P-S) \right\rbrace $, there are five equilibria: $(0,0)$, $(0,1)$, $(1,0)$, $(1,1)$, and $\left(\frac{1}{1+\theta},\frac{1}{2} \right) $.
Subsequently, we will derive several theorems regarding the stability of these equilibria.

\subsection{The System \eqref{rep_eq} Has Five Equilibrium Points.}\label{sub3.1}
When either of the conditions is satisfied:  $T-R>{\rm max}\left\lbrace\frac{P-S}{1-k},(1-k)(P-S) \right\rbrace $ or $T-R<{\rm min}\left\lbrace\frac{P-S}{1-k},(1-k)(P-S) \right\rbrace $, the system has five equilibria and exhibits two distinct dynamics.
These dynamics are primarily determined by whether the interior equilibrium is a center or a saddle point.
The specific conclusions are provided in the following theorems.

\begin{theorem}\label{thm1}
	If $T-R>{\rm max}\left\lbrace\frac{P-S}{1-k},(1-k)(P-S) \right\rbrace $ is satisfied, all corner equilibrium points are unstable saddle points, and the interior equilibrium point $\left(\frac{1}{1+\theta},\frac{1}{2}\right)$ is a center. Furthermore, the trajectory of each solution of the system \eqref{rep_eq} is a closed orbit in $D=(0,1)\times(0,1)\backslash\left\lbrace\left(\frac{1}{1+\theta},\frac{1}{2}\right) \right\rbrace $.
\end{theorem}

The detailed proof of Theorem \ref{thm1} can be found in \ref{A.2}.
This theorem indicates that if the condition $T-R>{\rm max}\left\lbrace\frac{P-S}{1-k},(1-k)(P-S) \right\rbrace $ is satisfied, the frequency of cooperation and the environmental state will exhibit persistent oscillations over time, regardless of the system's initial state. This indicates that  the system will consistently exhibit self-recovery capabilities.

In the following, a proposition concerning the existence of a heteroclinic cycle is presented under the conditions of Theorem \ref{thm1}.
\begin{proposition}\label{cor1}
	If $T-R>{\rm max}\left\lbrace\frac{P-S}{1-k},(1-k)(P-S) \right\rbrace $ is satisfied, there exists a heteroclinic cycle, represented by $(0,0)\to (1,0)\to (1,1)\to (0,1)\to (0,0)$.
\end{proposition}

The proof of Proposition \ref{cor1} is provided in \ref{A.2}. Based on Theorem \ref{thm1} and Proposition \ref{cor1}, it can be observed that all solutions of the system \eqref{rep_eq} located inside the heteroclinic cycle are periodic, excluding the interior equilibrium $\left(\frac{1}{1+\theta},\frac{1}{2}\right)$.

However, periodic solutions may not always exist when there are five equilibria.
When $T-R<{\rm min}\left\lbrace\frac{P-S}{1-k},(1-k)(P-S) \right\rbrace $, a phenomenon occurs in which the cooperation or defection disappears, while the environment tends towards a state of saturation or depletion, respectively. The corresponding conclusions are provided in the theorem below.

\begin{theorem}\label{thm2}
	If $T-R<{\rm min}\left\lbrace\frac{P-S}{1-k},(1-k)(P-S) \right\rbrace $ is satisfied, the interior equilibrium point $\left(\frac{1}{1+\theta},\frac{1}{2}\right)$ is an unstable saddle point. Besides, the equilibrium points $ (0, 0) $ and $ (1, 1) $ are stable nodes, while $ (0, 1) $ and $ (1, 0) $ are unstable nodes.
\end{theorem}

The proof details of Theorem \ref{thm2} are given in \ref{A.2}.
When $T-R<{\rm min}\left\lbrace\frac{P-S}{1-k},(1-k)(P-S) \right\rbrace $, the system \eqref{rep_eq} exhibits bistability, with two stable equilibria represented by $ (0, 0) $ and $ (1, 1) $, respectively. Since the saddle point $ (\frac{1}{1+\theta}, \frac{1}{2}) $ is the sole interior equilibrium, the basins of attraction of $ (0, 0) $ and $ (1, 1) $ are bounded by the stable manifold of $ (\frac{1}{1+\theta}, \frac{1}{2}) $.
Accordingly, the global dynamics correspond to that for any initial condition $(x_0,m_0)$ where $x_0\in(0,1)$ and $m_0\in(0,1)$ and the point does not lie on the stable manifolds of the saddle point,
the system will inevitably converge to one of two states: either cooperators dominate in the most abundant environmental state, or defectors dominate in a depleted environmental state.
Moreover, a higher frequency of cooperation or a more abundant environmental state is conducive to the system tending toward a more favorable stable state.

In the next subsection, we will present the results of the stability analysis for cases where the system has seven equilibrium points.

\subsection{The System \eqref{rep_eq} Has Seven Equilibrium Points.}

When $1-k<\frac{T-R}{P-S}<\frac{1}{1-k}$, the system \eqref{rep_eq} has seven equilibria, resulting in more complicated dynamical behaviors.
Compared to the case with five equilibria, there are two more equilibria, $(x^{*},0)$ and $(x^{*},1)$. They represent the coexistence of cooperators and defectors in the depleted and replete environment, respectively.
Correspondingly, the specific conclusions about the system dynamics are given by the following theorems.

\begin{theorem}\label{thm3}
	If $ \frac{\theta k-\theta+1}{1-\theta-k}(P-S)<T-R< {\rm max}\left\lbrace\frac{P-S}{1-k},(1-k)(P-S) \right\rbrace$ is satisfied, the interior equilibrium point $\left(\frac{1}{1+\theta},\frac{1}{2}\right)$ for the system \eqref{rep_eq} is a center.
	\begin{enumerate}[(1)]
		\item
		When $T-R>0$, $P-S<0$, and $1-k<\frac{T-R}{P-S}<\frac{\theta k-\theta+1}{1-\theta-k}$, the equilibrium point $(0,0)$ is a stable node; $ (0, 1) $ is an unstable node; $\left(x^{*},1\right)$, $\left( x^{*},0\right)$, $(1,0)$,\vphantom{$\frac{k-1}{k-2}$} and $(1,1)$ are unstable saddle points.
		\item
		When $T-R<0$, $P-S>0$, and $\frac{\theta k-\theta+1}{1-\theta-k}<\frac{T-R}{P-S}<\frac{1}{1-k}$, the equilibrium point $(1,1)$ is a stable node; $ (1, 0) $ is an unstable node; $\left( x^{*},1\right)$, $\left( x^{*},0\right)$, $(0,0)$,\vphantom{$\frac{k-1}{k-2}$} and $(0,1)$ are unstable saddle points.
	\end{enumerate}
\end{theorem}

The proof of Theorem \ref{thm3} is provided in \ref{A.4}.
This theorem indicates that the system \eqref{rep_eq} exhibits a coexistence of persistent oscillations and the domination of cooperation or defection  when the inequality $ \frac{\theta k-\theta+1}{1-\theta-k}(P-S)<T-R< {\rm max}\left\lbrace\frac{P-S}{1-k},(1-k)(P-S) \right\rbrace$ is satisfied.
More specifically, the interior equilibrium point $\left(\frac{1}{1+\theta},\frac{1}{2}\right)$ is a center, similar to the result in Theorem \ref{thm1}, and is surrounded by periodic solutions. Additionally, the system has a stable equilibrium at either $(0,0)$, representing that defectors dominate in a depleted environment, or $(1,1)$, representing that cooperators dominate in a replete environment.

Similar to Proposition \ref{cor1}, a heteroclinic cycle can also be obtained under the conditions of Theorem \ref{thm3}.

\begin{proposition}\label{cor2}
	If $ \frac{\theta k-\theta+1}{1-\theta-k}(P-S)<T-R< {\rm max}\left\lbrace\frac{P-S}{1-k},(1-k)(P-S) \right\rbrace$ is satisfied, there exists a heteroclinic cycle. Specifically,
	\begin{enumerate}[(1)]
		\item when $T-R>0$, $P-S<0$, and $1-k<\frac{T-R}{P-S}<\frac{\theta k-\theta+1}{1-\theta-k}$, the heteroclinic cycle is $(1,0)\to (1,1)\to \left( x^{*},1\right)\to\left( x^{*},0\right)\to(1,0)$\vphantom{$\frac{k-1}{k-2}$};
		\item when $T-R<0$, $P-S>0$, and $\frac{\theta k-\theta+1}{1-\theta-k}<\frac{T-R}{P-S}<\frac{1}{1-k}$, the heteroclinic cycle is  $ (0,0) \to\left(x^{*},0\right)\to\left(x^{*},1\right)\to(0,1)\to(0,0)$\vphantom{$\frac{k-1}{k-2}$}.
	\end{enumerate}
\end{proposition}

The proof of Proposition \ref{cor2} is provided in \ref{A.4}.
Indeed, all trajectories starting from $(x_0,m_0)$ within the interior of the heteroclinic cycle, but not at the center, form closed periodic orbits. In other words, inside the heterocyclic cycle, both  the frequency of cooperation and the environmental state display persistent oscillations.
By combining Theorem \ref{thm3} and Proposition \ref{cor2}, it is obtained that
when $T-R>0$, $P-S<0$, and $1-k<\frac{T-R}{P-S}<\frac{\theta k-\theta+1}{1-\theta-k}$, the system  initialized with a low frequency of cooperators will inevitably reach a state where defectors dominate in the depleted environmental state, even if its initial environment state is favorable.
Conversely, when $T-R<0$, $P-S>0$, and $\frac{\theta k-\theta+1}{1-\theta-k}<\frac{T-R}{P-S}<\frac{1}{1-k}$, the system initialized with a high frequency of cooperators will converge to a state where cooperators dominate in the replete environment, regardless of the initial environmental state.
Furthermore, as indicated by Proposition \ref{cor1} and \ref{cor2}, a heteroclinic cycle exists if and only if  the system's interior equilibrium point $\left(\frac{1}{1+\theta},\frac{1}{2}\right)$ is a center.

It is worth noting that due to the existence of a heteroclinic cycle, it is clear to observe the global dynamics of the system with oscillatory behavior and easier to analyze the impact of population structure on the system dynamics.
For instance, when $T-R > 0$ and $ P-S > 0$, the frequency of cooperators and the  environmental state oscillate cyclically and indefinitely, regardless of the number of neighbors $k$.
However, when $ \frac{\theta k-\theta+1}{1-\theta-k}(P-S)<T-R< {\rm max}\left\lbrace\frac{P-S}{1-k},(1-k)(P-S) \right\rbrace$, the system exhibits two distinct behaviors: oscillations within the heteroclinic cycle and dominance of cooperation or defection outside of it.
Therefore, the size of the regions where these two dynamics occur separately is influenced by the number of neighbors. Specifically, the number of neighbors impacts the value of \(x^{*}\), thereby determining the size of the heteroclinic cycle.

We also find an alternative dynamic of the system \eqref{rep_eq} that includes seven equilibrium points as described in the following theorem.

\begin{theorem}\label{thm4}
	If ${\rm min}\left\lbrace\frac{P-S}{1-k},(1-k)(P-S) \right\rbrace <T-R< \frac{\theta k-\theta+1}{1-\theta-k}(P-S)$ is satisfied, the interior equilibrium point $\left(\frac{1}{1+\theta},\frac{1}{2}\right)$ is a saddle point. Besides,
	\begin{enumerate}[(1)]
		\item when $T-R>0$, $P-S<0$, and $ \frac{\theta k-\theta+1}{1-\theta-k}<\frac{T-R}{P-S}<\frac{1}{1-k} $, $\left( x^{*},1\right)$ and $(0,0)$ are stable nodes; $(0,1)$ and $\left( x^{*},0\right)$ are unstable nodes; $(1, 0) $ and $ (1, 1) $ are unstable saddle points\vphantom{$\frac{k-1}{k-2}$};
		\item when $T-R<0$, $P-S>0$, and $1-k <\frac{T-R}{P-S}<\frac{\theta k-\theta+1}{1-\theta-k} $, $\left( x^{*},0\right)$ and $(1,1)$ are stable nodes; $(1,0)$ and $\left( x^{*},1\right)$ are unstable nodes; $ (0, 0) $ and $ (0, 1) $ are unstable saddle points\vphantom{$\frac{k-1}{k-2}$}.
	\end{enumerate}
\end{theorem}

The proof details of Theorem \ref{thm4} are given in \ref{A.5}.
This theorem suggests that when the inequality ${\rm min}\left\lbrace\frac{P-S}{1-k},(1-k)(P-S) \right\rbrace <T-R< \frac{\theta k-\theta+1}{1-\theta-k}(P-S)$ is satisfied, there exist two stable equilibria: a stable coexistence of cooperation and defection, and a dominant of defection or cooperation.
There are two stable manifolds converging to the interior saddle point, each originating from a separate unstable node.
When $T-R>0$, $ P-S<0 $, and $ \frac{\theta k-\theta+1}{1-\theta-k}<\frac{T-R}{P-S}<\frac{1}{1-k} $,  the coexisting stable equilibrium is $ (x^{*}, 1) $, and the stable dominance equilibrium is $ (0,0) $.
If the initial condition $(x_0,m_0)$ lies below the two stable manifolds originating from $(0,1)$ and $(x^{*},0)$, the solution will converge to $(0,0)$. Otherwise, it will converge to $(x^*,1)$.
In other words, if the initial cooperative frequency is low or the environmental state is poor, the system will evolve to a state where defectors dominate in a depleted environment. Otherwise, cooperators and defectors will coexist, and the environmental state will remain replete.
When $T-R<0$, $  P-S>0 $, and $1-k <\frac{T-R}{P-S}<\frac{\theta k-\theta+1}{1-\theta-k} $, the stable coexisting equilibrium of the system is $ (x^{*}, 0) $, and the stable dominance equilibrium is $ (1,1) $.
In this case, if the initial cooperative frequency is high or the environmental state is rich, the system will eventually evolve into a state where  cooperators dominate in the replete environmental state.

In summary, the coevolutionary dynamics of feedback-evolving games in structured populations are more complicated compared to those in well-mixed populations (i.e., in the limit case of $k\to\infty$). This fully reflects the significant impact of the number of neighbors on the coevolutionary dynamics.
Moreover, according to the theoretical results obtained above, it is found that,
$\theta$, a parameter representing the ratio of enhancement rates to degradation rates for cooperators and defectors, does not alter the diversity of dynamics in the system.

\section{Numerical Results}\label{example}

In this section, we will present numerical examples to verify all the possible coevolutionary dynamics discussed above and analyze the impact of the number of neighbors on these dynamics.

\subsection{Numerical Results for the Case of Five Equilibrium Points.}
If not specified, set $\epsilon=0.1$, $\theta=1.8$ and $k=3$. Under these conditions, the interior equilibrium of the system \eqref{rep_eq} is $(\frac{5}{14},\frac{1}{2})$.

\begin{figure}[thbp!]
\centering
\begin{overpic}[width=0.8\textwidth]{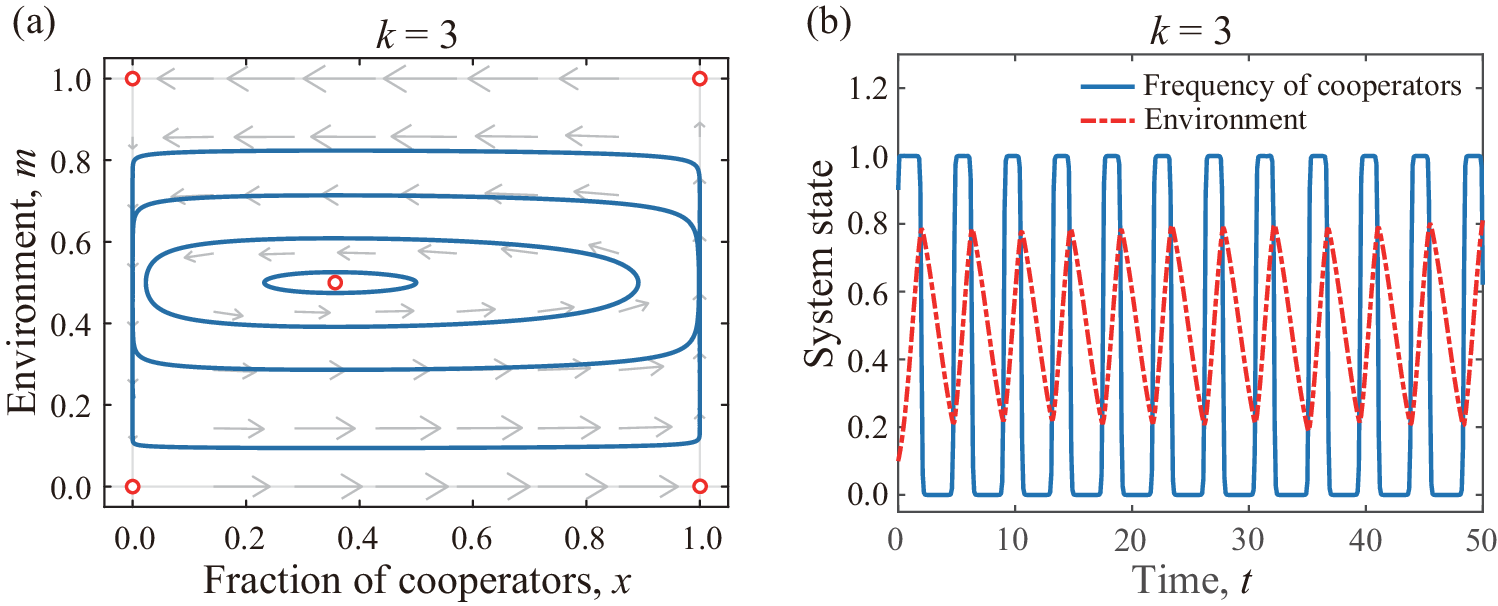}
\end{overpic}
	\caption{Persistent oscillations of feedback-evolving games in structured populations when $T-R>0$ and $P-S>0$.
		Panel~(a) displays the phase portrait of the system. The blue curves denote trajectories of some solutions. Open red dots denote non-asymptotically stable equilibria of the system; Panel~(b) depicts the time series plot with initial conditions $x(0)=0.9$ and $m(0)=0.1$. Parameters: $\epsilon=0.1$, $\theta=1.8$, $k=3$, $R=3$, $S=0$, $T=5$, and $P=1$.}
	\label{fig_1}
\end{figure}

To verify the results of Theorem \ref{thm1} and Proposition \ref{cor1}, we provide a numerical example in Fig.~\ref{fig_1}.
For the example, we assume $R=3$, $S=0$, $T=5$, and $P=1$, which satisfy the conditions $T-R>0$ and $P-S>0$.
In Fig.~\ref{fig_1}(a), the vector fields on the phase plane are depicted.
We observe that trajectories along the boundary of the region form a neutral heteroclinic cycle. Additionally, any trajectories within this cycle, except at the point $(\frac{5}{14}, \frac{1}{2})$,  represents a closed orbit.
Furthermore, Fig.~\ref{fig_1}(b) shows a time series plot about the fraction of cooperators and the environmental state. We further verify that the solution, starting from the initial condition, completes one full counterclockwise revolution and returns to its starting point.
This indicates that the frequency of cooperators and the environmental state exhibit persistent oscillations between 0 and 1, which further verifies Theorem \ref{thm1}.

\begin{figure}[thbp!]
	\centering
	\begin{overpic}[width=1\textwidth]{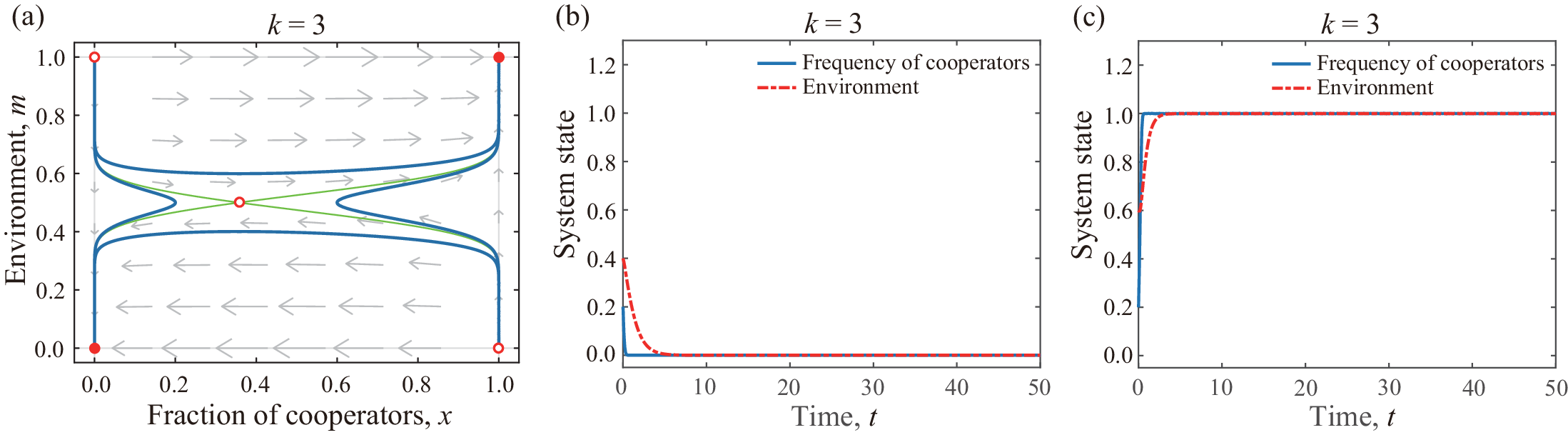}
	\end{overpic}
	\caption{Bistability for feedback-evolving games in structured populations when $T-R<0$ and $P-S<0$.
		Panel~(a) depicts the phase portrait. Green line indicates the stable or unstable manifold of the interior saddle point. Open (closed) solid red dots indicate unstable (stable) equilibrium points, respectively. Panel~(b) and (c) depict the time series plot with initial conditions $(0.2,0.4)$ and $(0.2,0.6)$, respectively. Parameters: $\epsilon=0.1$, $\theta=1.8$, $k=3$, $R=5$, $S=1$, $T=3$, and $P=0$.}
	\label{fig_2}
\end{figure}

To verify the result of Theorem \ref{thm2}, we provide an example in Fig.~\ref{fig_2}. Assume that $R=3$, $S=0$, $T=5$, and $P=1$, which satisfy the conditions $T-R<0$ and $P-S<0$.
In Fig.~\ref{fig_2}(a), the phase plane is depicted with two green curves.
One of the curves connects the points $ (0,1) $ and $ (1,0) $, representing two stable manifolds of the saddle point $(\frac{5}{14},\frac{1}{2})$, with the vector field directed toward the saddle point. The other green curve connects the points $ (0,0) $ and $ (1,1) $, representing two unstable manifolds, with the vector field pointing away from the saddle point.
Clearly, solutions above the stable manifolds eventually converge towards the stable equilibrium $(1,1)$, representing a state where cooperators dominate in the most abundant environmental state. Conversely,  solutions below the unstable manifolds converge to the stable equilibrium $(0,0)$, indicating a state where  defectors dominate in the most scarce environmental state.
The time series plots in Fig.~\ref{fig_2}(b) and (c) confirm the evolutionary outcomes of the system, where either cooperators dominate in a favorable  environmental state or defectors prevail in an unfavorable environmental state.
Therefore the result of Theorem \ref{thm2} is verified.

\subsection{Numerical Results for the Case of Seven Equilibrium Points.}
Here, we will verify the results of the coevolutionary dynamics when the system has seven equilibria.
As usual, set $\epsilon=0.1$, $\theta=1.8$, and $k=3$. Then, the interior equilibrium of the system  \eqref{rep_eq} is $(\frac{5}{14},\frac{1}{2})$.
\begin{figure}[thbp!]
	\centering
	\begin{overpic}[width=1\textwidth]{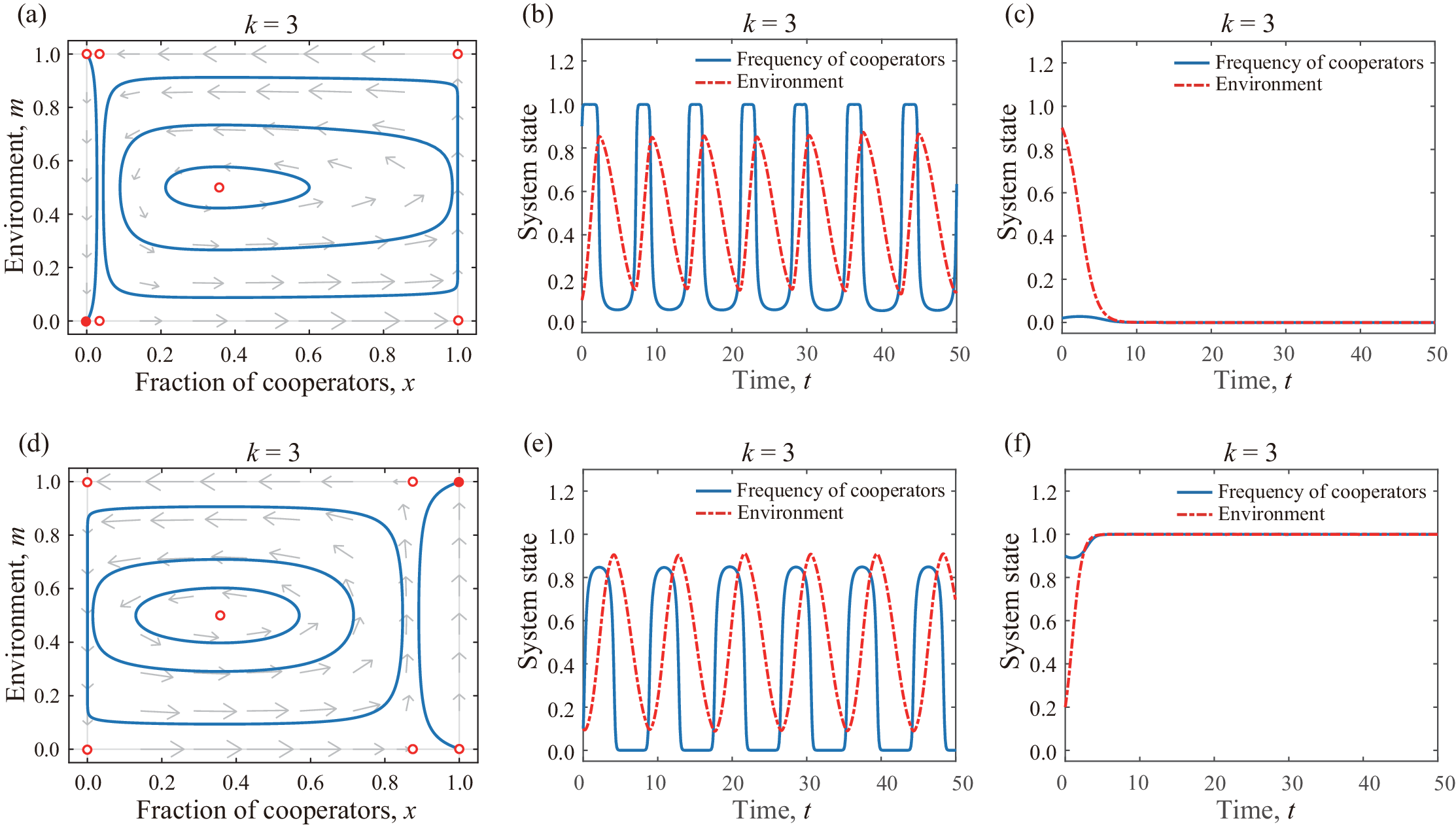}
	\end{overpic}
	\caption{Coexistence of oscillation and dominance for feedback-evolving games in structured populations when $ \frac{\theta k-\theta+1}{1-\theta-k}(P-S)<T-R< {\rm max}\left\lbrace\frac{P-S}{1-k},(1-k)(P-S) \right\rbrace$.
		There exist seven equilibria, and stable equilibria  are depicted with closed red dots, while unstable equilibria  are depicted with open red dots. Panels~(a) and (d) show the phase portraits where blue lines denote the trajectories of some solutions. In the top panels, panels~(b) and (c) depict the time course of the frequencies of cooperators (blue solid lines) and the state of the environment (red dashed lines), initialized at $ (0.9,0.1)$ and $ (0.02,0.9)$, respectively.
		In the bottom panels, panels~(e) and (f) depict the time course of the frequencies of cooperators (blue solid lines) and the state of the environment (red dashed lines), initialized at $(0.1,0.1)$ and $(0.9,0.2)$, respectively.
		Parameters in panels~(a)-(c): $\epsilon=0.1$,  $\theta=1.8$, $k=3$, $R=3$, $S=1$, $T=4.9$, and $P=0$.
		Parameters in panels~(d)-(f): $\epsilon=0.1$,  $\theta=1.8$, $k=3$, $R=3.6$, $S=0$, $T=3$, and $P=1$.}
	\label{fig_1'}
\end{figure}

To verify Theorem \ref{thm3}, we consider the parameter values as $R=3$, $S=1$, $T=4.9$, and $P=0$, which satisfy $T-R>0$, $P-S<0$, and $1-k<\frac{T-R}{P-S}<\frac{\theta k-\theta+1}{1-\theta-k}$. As shown in Fig.~\ref{fig_1'}(a)-(c), the system exhibits a single stable equilibrium at $(0,0)$, corresponding to a state where defectors dominate in a depleted environment.
Next, we set $R=3.6$, $S=0$, $T=3$, and $P=1$, which satisfy $T-R<0$, $P-S>0$, and $\frac{\theta k-\theta+1}{1-\theta-k}<\frac{T-R}{P-S}<\frac{1}{1-k}$. As depicted in Fig.~\ref{fig_1'}(d)-(f), the system has a single  stable equilibrium $(1,1)$,  representing a state where cooperators dominate in a replete environmental state. In both cases, if the initial conditions lie within a heteroclinic cycle, the frequency of cooperators and the environmental state exhibit persistent oscillations. The heteroclinic cycle is characterized by the trajectories \( (1,0) \to (1,1) \to (\frac{1}{29},1) \to (\frac{1}{29},0) \to (1,0) \) and \( (0,0) \to (\frac{7}{8},0) \to (\frac{7}{8},1) \to (0,1) \to (0,0) \), respectively.
Therefore, the system demonstrates a coexistence of oscillation and dominance when $ \frac{\theta k-\theta+1}{1-\theta-k}(P-S)<T-R< {\rm max}\left\lbrace\frac{P-S}{1-k},(1-k)(P-S) \right\rbrace$. This verifies the result of Theorem \ref{thm3}.

\begin{figure}[thbp!]
	\centering
	\begin{overpic}[width=1\textwidth]{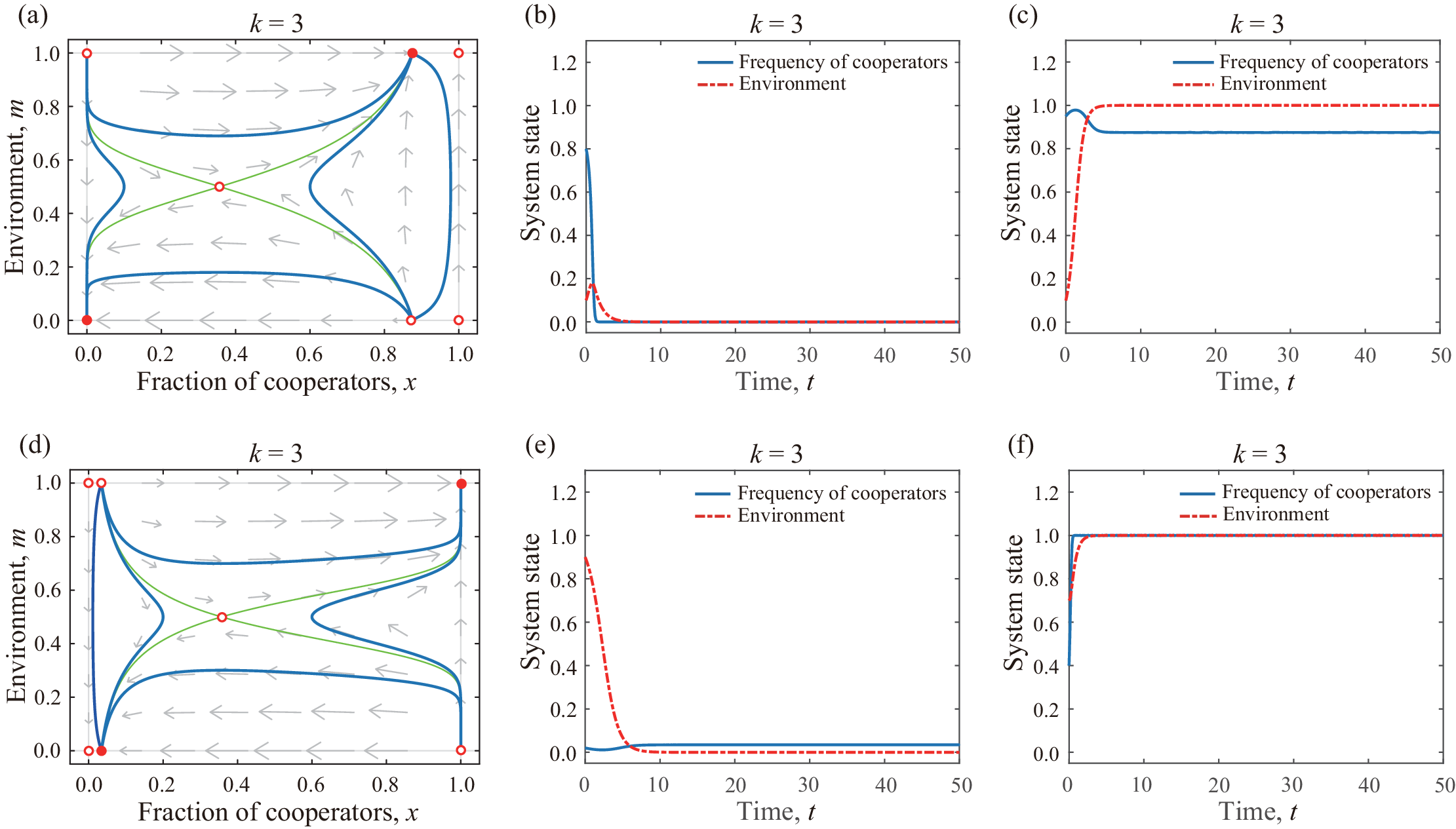}
	\end{overpic}
	\caption{Coexistence of cooperation and defection for feedback-evolving games in structured populations when ${\rm min}\left\lbrace\frac{P-S}{1-k},(1-k)(P-S) \right\rbrace <T-R< \frac{\theta k-\theta+1}{1-\theta-k}(P-S)$. There are seven equilibria, and stable equilibria  are depicted with closed red dots, while unstable equilibria  are depicted with open red dots. Panels~(a) and (d) depict the phase diagram with some trajectories. In the top panels, panels~(b) and (c) show the time course of the frequencies of cooperators (blue solid line) and the state of the environment (red dash line) for initial conditions  $(x(0),m(0))=(0.8,0.1)$ and $(x(0),m(0))=(0.95,0.1)$, respectively. In the bottom panels, panels~(e) and (f) show the time course of the frequencies of cooperators (blue solid line) and the state of the environment (red dash line) for initial conditions $(x(0),m(0))=(0.02,0.9)$ and $(x(0),m(0))=(0.4,0.7)$, respectively. Parameters in panels~(a)-(c): $\epsilon=0.1$,  $\theta=1.8$, $k=3$, $R=3$, $S=1$, $T=3.6$, and $P=0$.	Parameters in panels~(d)-(f): $\epsilon=0.1$,  $\theta=1.8$, $k=3$, $R=4.9$, $S=0$, $T=3$, and $P=1$.}
	\label{fig_2'}
\end{figure}
To validate Theorem \ref{thm4}, we consider the parameter values as $R=3$, $S=1$, $T=3.6$, and $P=0$, which satisfy  $T-R>0$, $P-S<0$, and $\frac{\theta k-\theta+1}{1-\theta-k}<\frac{T-R}{P-S}<\frac{1}{1-k}$.
As shown in the top row of Fig.~\ref{fig_2'}, the system has two stable equilibria, $(0,0)$ and $(\frac{7}{8},1)$.
This indicates that, regardless of the initial conditions, the environmental state will evolve toward one of two extreme states: depletion or repletion.  Furthermore, when the system reaches a replete environmental state, the majority of individuals tend to cooperate.
Next, we set $R=4.9$, $S=0$, $T=3$, and $P=1$, which satisfy $T-R<0$, $P-S>0$, and $1-k <\frac{T-R}{P-S}<\frac{\theta k-\theta+1}{1-\theta-k} $. As shown in the bottom row of Fig.~\ref{fig_2'}, the system also has two stable equilibria, $(\frac{1}{29},0)$ and $(1,1)$. In this case, a similar result is observed: the system evolves towards either a depleted or replete environmental state, with full cooperation occurring in the replete environmental state. Thus, the system exhibits a coexistence of cooperation and defection when ${\rm min}\left\lbrace\frac{P-S}{1-k},(1-k)(P-S) \right\rbrace <T-R< \frac{\theta k-\theta+1}{1-\theta-k}(P-S)$. This confirms Theorem \ref{thm4}.

\subsection{Diversity of Coevolutionary Dynamics in Structured Populations.}

\begin{figure}[thbp!]
	\centering
	\begin{overpic}[width=1\textwidth]{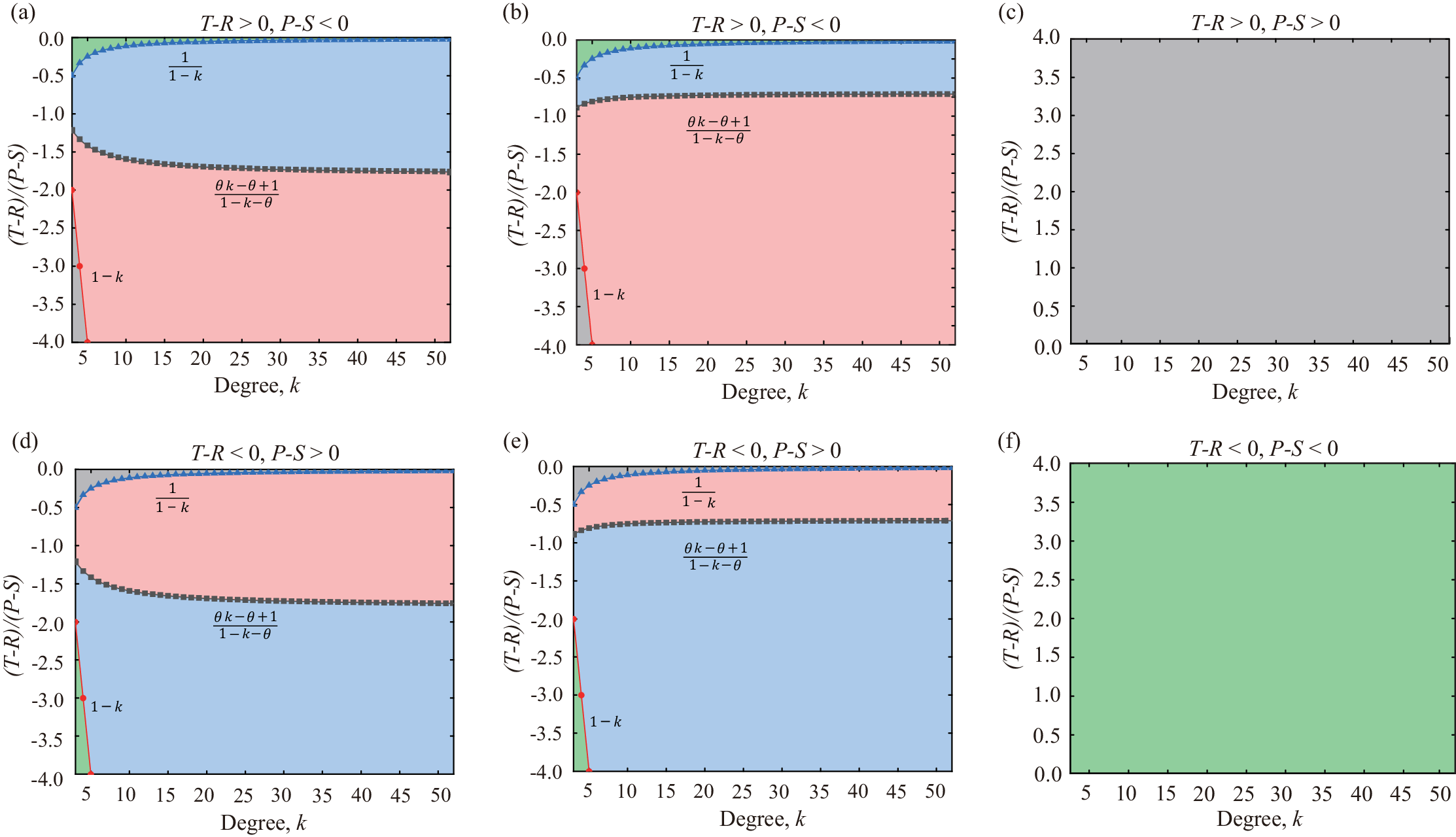}
	\end{overpic}
	\caption{Schematic diagrams illustrating the diversity of coevolutionary dynamics for feedback-evolving games in structured populations.
		Set $\epsilon = 0.1$. First column (Middle column) displays the coevolutionary dynamics of the population with $\theta = 1.8$ ($\theta = 0.7$) under the conditions $T - R > 0$, $P - S < 0$ and $T - R < 0$, $P - S > 0$, respectively.  Panels~(c) and (f) depict the evolutionary dynamics for any $\theta > 0$ under the conditions $T - R > 0$, $P - S > 0$, and $T - R < 0$, $P - S < 0$, respectively. The dynamics in the gray, green, red, and blue regions correspond to oscillations, bistability, the coexistence of oscillation and domination, and the coexistence of cooperation and defection, respectively.}
	\label{fig_6}
\end{figure}

In order to help readers easily overview the evolutionary outcomes of our dynamical system for different parameter regions, we provide several schematic plots to elucidate the diversity of coevolutionary dynamics in structured populations. As shown in Fig.~\ref{fig_6}, different colors correspond to the various dynamics previously described.
Specifically, gray regions  indicate oscillations, green regions  indicate bistability, red regions represent the coexistence of oscillation and domination, and blue regions mean the coexistence of cooperation and defection. As the degree $ k $ approaches infinity, the population tends to become well-mixed, exhibiting only two dynamical behaviors when $ \frac{T-R}{P-S} <0$: the coexistence of oscillation and domination, and the coexistence of cooperation and defection. This reveals that the dynamics in a structured population are more diverse and complicated than those in a well-mixed population, and this disparity on dynamics becomes more pronounced as the parameter $k$ decreases. Furthermore, comparing the first and second columns of Fig.~\ref{fig_6}, we find that the value of $\theta$ does not impact the diversity of the dynamics, but it can influence their distribution.

In addition, we can learn that the population structure plays a crucial role in shaping coevolutionary dynamics. In particular, the number of neighbors is a key parameter in the spatially structured population. Subsequently, we will investigate how the neighbor number influences the coevolutionary dynamics, particularly its effect on the evolution of the environmental state.

\subsection{Numerical Results for the Impacts of neighborhood size on the Coevolutionary Dynamics.}

\begin{figure}[thbp!]\centering
	\begin{overpic}[height=5cm]{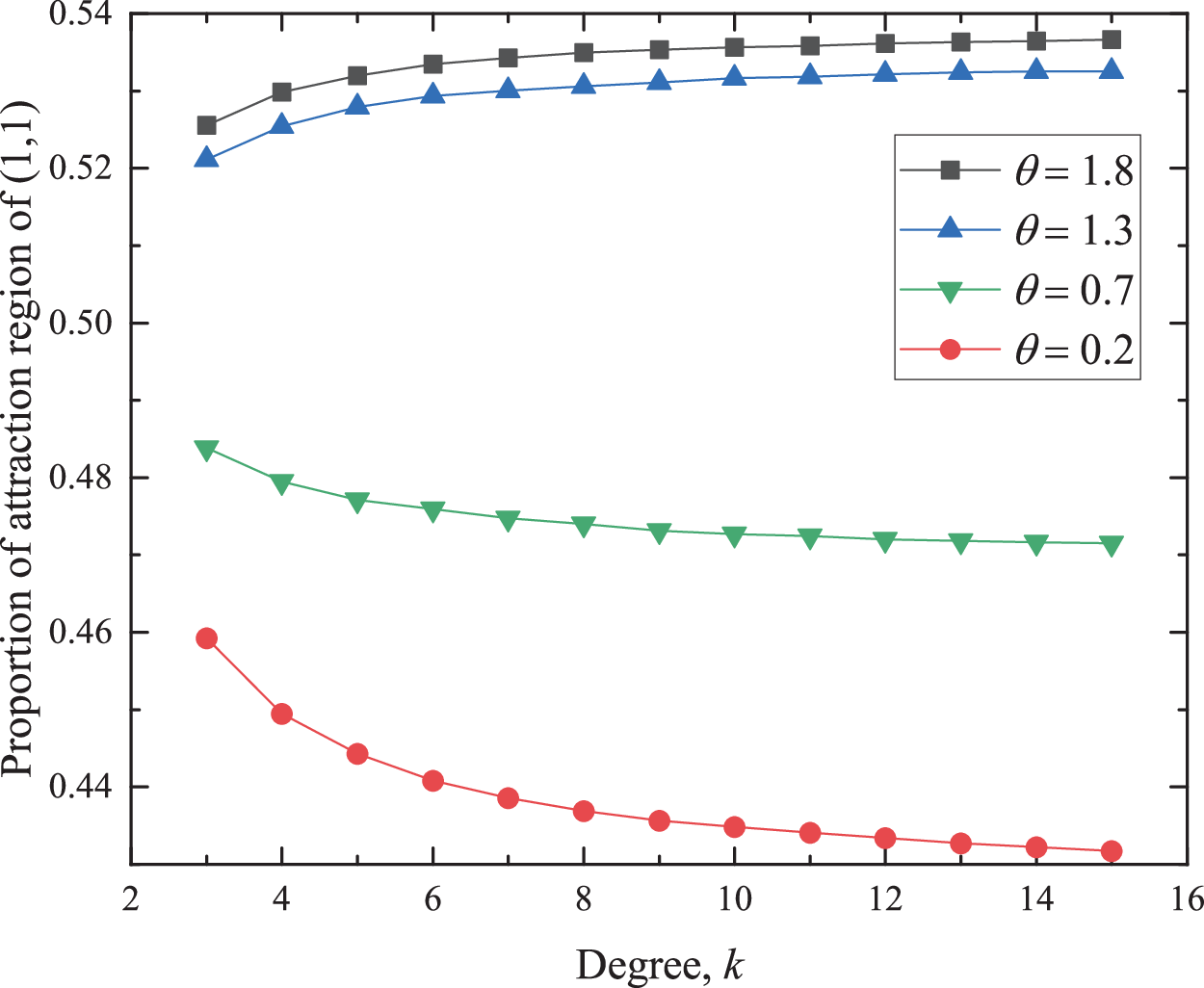}
	\end{overpic}
	\caption{Impact of the degree $k$ on the attraction region of the attractor with sufficient resources in the dynamics of bistability. There are two stable attractors, $(0,0)$ and $(1,1)$. Using Monte Carlo simulations, the panel shows the change in the basin of attraction of $(1,1)$ with increasing $k$ for different values of $\theta$. It  indicates that for a larger $\theta$, the attraction region expands as the number of neighbors increases, whereas for a smaller $\theta$, it shrinks with increasing $k$.
		Parameters: $\epsilon=0.1$, $R=5$, $S=1$, $T=3$, and $P=0$.}
	\label{fig7_1}
\end{figure}

\begin{figure}[thbp!]\centering
	\begin{overpic}[width=0.8\textwidth]{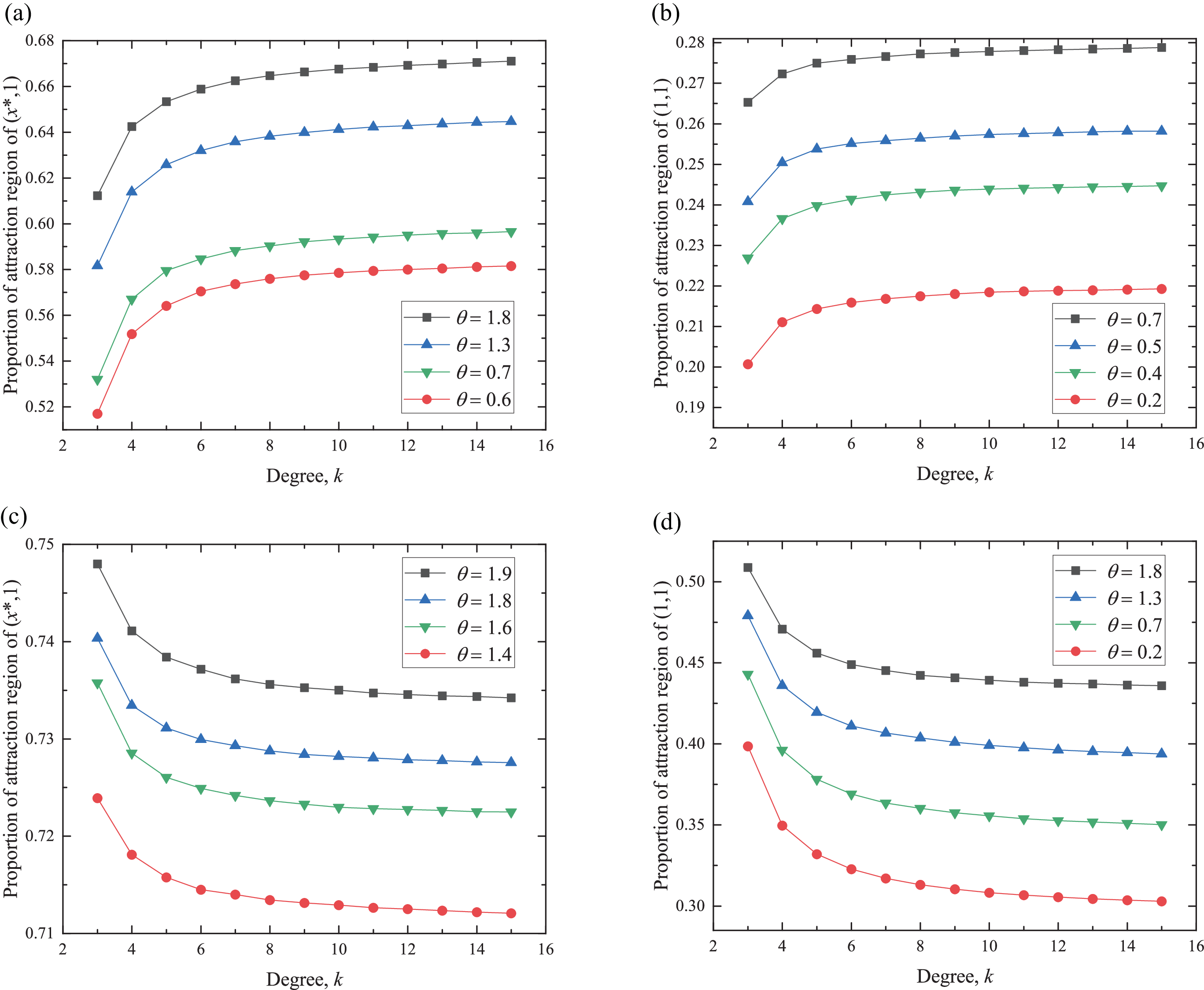}
	\end{overpic}
	\caption{Impact of the degree $k$ on the attraction region of the attractor with sufficient resources in the coexistence of cooperation and defection.
		First column depicts the size of the attraction region of $(x^{*},1)$ for $\frac{T-R}{P-S}=-\frac{3}{5}$ and $\frac{T-R}{P-S}=-\frac{11}{10}$, respectively.
		Second column depicts the size of the attraction region of $(1,1)$ for $\frac{T-R}{P-S}=-\frac{9}{10}$ and $\frac{T-R}{P-S}=-\frac{19}{10}$, respectively.
		When $\frac{T-R}{P-S}$ is relatively high, for a proper value of $\theta$ that does not affect the overall dynamics, the size of the attraction region corresponding to the attractor with sufficient resources increases as the value of $ k $ increases (panels~(a) and (b)). However, when  $\frac{T-R}{P-S}$ is relatively low, for a proper value of $\theta$ that does not affect the overall dynamics, the size of the attraction region corresponding to the attractor with sufficient resources decreases as the value of $ k $ increases (panels (c) and (d)).
		Parameters in panel~(a): $\epsilon=0.1$, $R=3$, $S=1$, $T=3.6$, and $P=0$.
		Parameters in panel~(b): $\epsilon=0.1$, $R=3.9$, $S=0$, $T=3$, and $P=1$.
		Parameters in panel~(c): $\epsilon=0.1$, $R=3$, $S=1$, $T=4.1$, and $P=0$.
		Parameters in panel~(d): $\epsilon=0.1$, $R=4.9$, $S=0$, $S=3$, and $P=1$.
	}
	\label{fig8}
\end{figure}

Note that the significant impact of neighborhood size $k$ on the system dynamics with fixed payoff parameters is complicated due to the diverse dynamics involved. Here, our primary focus is on the dynamics featuring two stable equilibria: bistability and the coexistence of cooperation and defection. Specifically, we analyze the basin of attraction of the equilibrium  characterized by the replete environmental state and its dependence on the number of neighbors. This analysis is crucial for understanding the maintenance of environmental state in a population.
The outcomes shown in Fig.~\ref{fig7_1} and Fig.~\ref{fig8} are obtained through Monte Carlo simulations.
To reach the expected accuracy for each value of the parameter $k$, we generate $10^6$ random points in the whole regions of $0\le x\le1$ and $0\le m\le1$ as initial states. The state trajectory starting from each initial state point is then computed by using numerical calculations and the system along each trajectory will converge to a specific equilibrium, either $(1, 1)$ or $(x^*, 1)$, where $x^*$ depends on the given value of parameter $k$. We finally compute the fraction of the trajectories which eventuate in the $(1, 1)$ state as the proportion of attraction region of the stable state $(1, 1)$, and the fraction of the trajectories which eventuate in the $( x^*, 1)$ state is evaluated as the ratio of attraction region of the stable state $( x^*, 1)$.

When the system exhibits bistability (corresponding to the case of $T-R<0$ and $P-S<0$), the impact of $k$ on dynamics is closely related to $\theta $.
There is an ideal stable state, $(1,1)$, where cooperators dominate in the replete environmental state.
As depicted in Fig.~\ref{fig7_1}, for higher values of $\theta$, a larger number of neighbors is more likely to be beneficial for environmental development. Conversely, a smaller number of neighbors tends to be more advantageous when the value of $\theta$ is lower.

When the system exhibits coexistence of cooperation and defection, we focus only on situations where the number and type of equilibria remain unchanged as $k$ varies. In this case, a favorable stable equilibrium,  $(x^{*},1)$ or $(1,1)$, is characterized by a replete environmental state.
Besides, the size of the attraction basin for this stable equilibrium is closely related to  $x^*$, which is independent of $\epsilon$. For simplicity, we assume $\epsilon=0.1$ without loss of generality.
As shown in Fig.~\ref{fig8}, we find that the role of $\theta$ in the correlation between the size of the attraction basin and the parameter $k$ can be negligible. Instead, the value of $\frac{T-R}{P-S}$, representing the ratio of temptation minus reward to punishment minus sucker's payoff, plays a pivotal role. Specifically, when the value of $\frac{T-R}{P-S}$ is relatively high, the attraction basin of the stable equilibrium $(x^{*},1)$ or $(1,1)$ expands as $k$ increases (see Fig.~\ref{fig8}(a) and (b)). Conversely, when the value of $\frac{T-R}{P-S}$ is relatively low, the attraction basin shrinks as $k$ increases (see Fig.~\ref{fig8}(c) and (d)). In other words, when the ratio of temptation minus reward to punishment minus sucker's payoff is closer to zero (but still negative), it is  beneficial to have more neighbors. However, when the ratio of temptation minus reward to punishment minus sucker's payoff is more negative, it is advantageous to have fewer neighbors.


\section{Conclusion and Discussion}\label{disc}

Individual behavior and the environmental state mutually influence each other, and their coevolution can be effectively described by the feedback-evolving game. Notably, population structure plays a pivotal role in shaping the interactions between individuals.
To more accurately capture the strategy-environment coevolutionary dynamics, we have introduced a model of feedback-evolving games in a structured population. Under varying parameter conditions, the system exhibits four distinct  types of dynamics: oscillation, bistability, the coexistence of oscillation and strategy dominance, and the stable coexistence of cooperation and defection.
We have provided some numerical examples to verify the presence of these evolutionary outcomes. Notably, coevolutionary dynamics in structured populations are more diverse and complicated compared to the results in well-mixed populations. Using Monte Carlo simulations, we have further explored the impact of population structure on the system's dynamics, particularly focusing on how the number of neighbors affects the attractive domain of the stable equilibrium characterized by the replete environmental state. In the case of bistable state, we have observed that when the ratio of cooperators' enhancement rate to defectors' degradation rate is higher, it tends to be more beneficial for environmental maintenance to have more neighbors. Conversely, it tends to be more beneficial to have fewer neighbors. In the case of the coexistence of cooperation and defection, the relationship between population structure and the environmental state is independent of the ratio of cooperators' enhancement rate to defectors' degradation rate, but is closely related to individual payoff parameters. Specifically, when the ratio of temptation minus reward to punishment minus sucker's payoff is larger, a larger number of neighbors is more beneficial for environmental maintenance. Conversely, if the ratio of temptation minus reward to punishment minus sucker's payoff is smaller, a fewer number of neighbors is preferable. Our results have demonstrated that population structure plays a pivotal role in the coevolutionary dynamics of feedback-evolving games. Specifically, the size of neighborhood significantly impacts environmental maintenance. 
Our findings provide valuable insights into preventing the tragedy of the commons, particularly in environments where individuals' strategic decisions directly impact shared resources. 

Essentially, our work is based on a symmetric form of payoff matrices, which simplifies the analysis of the coevolutionary dynamics. However, for more general payoff matrices, as discussed in Ref. \cite{Weitz_PNAS_2016}, the dynamics of feedback-evolving games in structured populations are expected to become more complicated, which is worthy of investigation. Therefore, future research could explore these more complex scenarios to deepen our understanding of the coevolutionary dynamics in feedback-evolving games. Additionally, within the framework of replicator dynamics in structured populations, there is significant potential to explore the coevolutionary dynamics with different updating rules \cite{Ohtsuki_Nature_2006}, such as death-birth and imitation updating. Investigating how these rules influence coevolutionary dynamics could provide deeper insights into feedback-evolving games. This line of research may also enhance our understanding of real-world systems, where individual behaviors and the environmental state intricately interact.
On another note, our model primarily considers linear effects. It is important to note that incorporating nonlinear dynamics may reveal additional insights, especially in situations where individual strategies exhibit disproportionately large or small effects under different conditions. For example, as population size or resource availability changes, non-linear interactions may better capture the complexity of real-world scenarios. For future work, we could study the impact of introducing nonlinear effects on the coevolutionary dynamics of  strategic actions and the environmental state.

\section*{Acknowledgment}
This research was supported by the National Natural Science Foundation of China (Grants No. 62473081 and No. 62036002), the Sichuan Science and Technology Program (Grant No. 2024NSFSC0436), and the National Research, Development and Innovation Office (NKFIH) under Grant No.~K142948.

\appendix
\section{The Stability Analysis of the Equilibrium Points}\label{Osci}

	\subsection{The Jacobian for each equilibrium point}\label{A.1}
	
	In Section \ref{result} of the main text, we have provided a classification of equilibrium points for the system \eqref{rep_eq}. The Jacobian for each equilibrium point is given by
	\begin{equation}\nonumber
		\begin{split}
			J(0,0)=&\begin{bmatrix}
				\frac{1}{\epsilon}\left(\delta_{PS}+\frac{\delta_{PS}+\delta_{TR}}{k-2} \right)  & 0\\
				0 & -1
			\end{bmatrix},
			\qquad
			J(0,1)=
			\begin{bmatrix}
				-\frac{1}{\epsilon}\left(\delta_{PS}+\frac{\delta_{PS}+\delta_{TR}}{k-2} \right) & 0\\
				0 & 1
			\end{bmatrix},\\
			J(1,0)=&
			\begin{bmatrix}
				-\frac{1}{\epsilon}\left(\delta_{TR}+\frac{\delta_{PS}+\delta_{TR}}{k-2} \right) & 0\\
				0 & \theta
			\end{bmatrix},
			\qquad
			J(1,1)=
			\begin{bmatrix}
				\frac{1}{\epsilon}\left(\delta_{TR}+\frac{\delta_{PS}+\delta_{TR}}{k-2} \right) & 0\\
				0 & -\theta
			\end{bmatrix},\\
			J(\frac{1}{1+\theta},\frac{1}{2})=&
			\begin{bmatrix}
				0 & -\frac{2\left(1-\frac{1}{\theta+1} \right)\left(\delta_{PS}+\frac{\delta_{TR}-\delta_{PS}}{\theta+1}+\frac{\delta_{PS}+\delta_{TR}}{k-2} \right)  }{\epsilon(\theta+1)}\\
				\frac{\theta+1}{4} & 0
			\end{bmatrix},
		\end{split}
	\end{equation}
	\begin{equation}\nonumber
		\begin{split}
			J\left( \frac{(1-k)\delta_{PS}-\delta_{TR}}{(k-2)(\delta_{TR}-\delta_{PS})},0\right)
			=\
			\begin{bmatrix}
				\frac{((k+1)\delta_{PS}+\delta_{TR})(\delta_{PS}+(k-1)\delta_{TR})}{-\epsilon(k-2)^{2}(\delta_{TR}-\delta_{PS})} & 0 \\
				0 & (\theta+1)\left( \frac{\delta_{TR}+(k-1)\delta_{PS}}{(2-k)(\delta_{TR}-\delta_{TR})}\right)-1
			\end{bmatrix},
		\end{split}
	\end{equation}
	and		
	\begin{equation}\nonumber
		\begin{split}
			J\left( \frac{(1-k)\delta_{PS}-\delta_{TR}}{(k-2)(\delta_{TR}-\delta_{PS})},1\right)
			=
			\begin{bmatrix}
				\frac{((k+1)\delta_{PS}+\delta_{TR})(\delta_{PS}+(k-1)\delta_{TR})}{\epsilon(k-2)^{2}(\delta_{TR}-\delta_{PS})} & 0\\
				0 & 1-(\theta+1)\left( \frac{\delta_{TR}+(k-1)\delta_{PS}}{(2-k)(\delta_{TR}-\delta_{PS})}\right)
			\end{bmatrix}.
		\end{split}
	\end{equation}
	
	Subsequently, we analyze the stability of the equilibrium points in various cases using the Lyapunov theory.

	\subsection{Proof of Theorem \ref{thm1}}\label{A.2} 
	
	When the inequality $T-R>{\rm max}\left\lbrace\frac{P-S}{1-k},(1-k)(P-S) \right\rbrace $ holds, the system \eqref{rep_eq} has five equilibrium points. Four of these are corner equilibrium points located at $(0,0)$, $(0,1)$, $(1,1)$, and $(1,0)$, while one is an interior equilibrium point at $\left(\frac{1}{1+\theta},\frac{1}{2}\right)$.
	Note that there are three cases included, which are (i) $T-R>0$, $P-S>0$; (ii) $ T-R>0$, $ P-S<0$, $ \frac{T-R}{P-S}<1-k$; (iii)  $ T-R<0 $, $ P-S>0 $, $ \frac{T-R}{P-S}>\frac{1}{1-k}$, respectively.
	For each case, we have
	$$
	\frac{1}{\epsilon}\left(\delta_{PS}+\frac{\delta_{PS}+\delta_{TR}}{k-2} \right)>0 \quad\quad{\rm and}\quad\quad  \frac{1}{\epsilon}\left(\delta_{TR}+\frac{\delta_{PS}+\delta_{TR}}{k-2} \right)>0,
	$$
	implying that the Jacobian of each corner equilibria has one positive eigenvalue and one negative eigenvalue.
	Thus all corner equilibrias are locally unstable saddle points. For the interior equilibrium point $\left(\frac{1}{1+\theta},\frac{1}{2}\right)$, the eigenvalues of its Jacibian are
	\begin{equation}\nonumber
		\begin{split}
			\lambda_{1}= \frac{\sqrt{(2k-4)\theta}}{\sqrt{\epsilon}(k-2)(1+\theta)}\sqrt{(\theta k+1-\theta)(\delta_{TR}-\delta_{PS})-(\theta k+k)\delta_{TR}}
		\end{split}
	\end{equation}
	and
	\begin{equation}\nonumber
		\begin{split}
			\lambda_{2}=
			-\frac{\sqrt{(2k-4)\theta}}{\sqrt{\epsilon}(k-2)(1+\theta)}\sqrt{(\theta k+1-\theta)(\delta_{TR}-\delta_{PS})-(\theta k+k)\delta_{TR}}\ ,
		\end{split}
	\end{equation} which are purely imaginary numbers with $k>2$.
	
	We set
	\begin{equation}\nonumber
		\begin{split}
			\Phi(x,m)=\ &\frac{1}{\epsilon}x(1-x)(1-2m)\left[\delta_{PS}+(\delta_{TR}-\delta_{PS})x+\frac{\delta_{PS}+\delta_{TR}}{k-2} \right]\\
			=\ &\frac{1}{\epsilon}x(1-x)(1-2m)[\phi x+\psi(k)]
		\end{split}
	\end{equation}
	and
	\begin{equation}\nonumber
		\Psi(x,m)=m(1-m)[\theta x-(1-x)]=m(1-m)[(\theta+1)x-1],
	\end{equation}
	where $\phi=\delta_{TR}-\delta_{PS}$ and $\psi(k)=\delta_{PS}+\frac{\delta_{PS}+\delta_{TR}}{k-2}$. It is obvious that $\Phi(x,m)$ and $\Psi(x,m)$ are analytic functions. Thus the interior equilibrium point $\left(\frac{1}{1+\theta},\frac{1}{2}\right)$ is a center or focus.
	
	Here, we search for a first integral $H(x,m)$ using the method of separation of variables.
	Consider
	$$
	V(x,m)=F(x)+G(m).
	$$
	The time derivation of $V(x,m)$ is given by
	\begin{equation}\nonumber
		\begin{split}
			\dot{V}_{t}(x,m)
			=\ & \frac{dF(x)}{dx}\dot{x}+\frac{dG(m)}{dm}\dot{m}\\
			=\ &
			\frac{dF(x)}{\epsilon dx}x(1-x)(1-2m)[\phi x+\psi(k)]+\frac{dG(m)}{dm}m(1-m)[(\theta+1)x-1].
		\end{split}
	\end{equation}We obtain $\dot{V}_{t}(x,m)=0$ provided
	\begin{equation}\nonumber
		\frac{dF(x)}{dx}\cdot\frac{x(1-x)[\phi x+\psi(k)]}{\epsilon[(\theta+1)x-1]}
		=
		\frac{dG(m)}{dm} \cdot \frac{m(1-m)}{2m-1}
		\equiv1.
	\end{equation}Then
	\begin{equation}\label{int1}
		\begin{split}
			\frac{dF(x)}{dx} &=\ \frac{\epsilon[(\theta+1)x-1]]}{x(1-x)[\phi x+\psi(k)]}
		\end{split}
	\end{equation}
	and
	\begin{equation}\label{int2}
		\begin{split}
			\frac{dG(m)}{dm} =\ \frac{2m-1}{m(1-m)}.
		\end{split}
	\end{equation}
	By integrating equations \eqref{int1} and \eqref{int2}, we obtain
	\begin{equation}\nonumber
		\begin{split}
			F(x)=&\frac{\epsilon}{(\phi+\psi(k))\psi(k)}\left[ ((\theta+1)\psi(k)+\phi)\ln(\phi x+\psi(k))-\theta\psi(k)\ln(1-x)
			\right.  \\  &\left.
			-(\phi+\psi(k))\ln x\right]+c_{1}\\
			=&
			\frac{\epsilon\theta}{\phi+\psi(k)}\ln\left( \frac{\phi x+\psi(k)}{1-x}\right) +\frac{\epsilon}{\psi(k)}\ln\left( \frac{\phi x+\psi(k)}{x}\right) +c_{1}
		\end{split}
	\end{equation}
	and
	$$G(m)=-\ln(m-m^{2})+c_{2},$$
	where $c_{1}$ and $c_{2}$ are constants.
	Then
	\begin{equation}\nonumber
		\begin{split}
			V(x,m)=&
			\frac{\epsilon\theta}{\phi+\psi(k)}\ln\left( \frac{\phi x+\psi(k)}{1-x}\right) +\frac{\epsilon}{\psi(k)}\ln\left( \frac{\phi x+\psi(k)}{x}\right) \\
			&-\ln(m-m^{2})+c_{3},
		\end{split}
	\end{equation}where $c_{3}$ is a constant. Let
	\begin{equation}\nonumber
		\begin{split}
			H(x,m)=&
			\frac{\epsilon}{(\phi+\psi(k))\psi(k)}\left[ ((\theta+1)\psi(k)+\phi)\ln(\phi x+\psi(k))-\theta\psi(k)\ln(1-x)\right.  \\  &\left.
			-(\phi+\psi(k))\ln x\right] -\ln(m-m^{2}),
		\end{split}
	\end{equation}
	which is a continuous first integral of the system. Hence, the interior equilibrium point $\left(\frac{1}{1+\theta},\frac{1}{2}\right)$ is a center, implying that solutions in the neighborhood of
	$\left(\frac{1}{1+\theta},\frac{1}{2}\right)$
	will form closed periodic orbits \cite{Hirsch_Aca_2012}.

	Actually, every solution of the system forms a closed orbit in $D=(0,1)\times(0,1)\backslash\left\lbrace\left(\frac{1}{1+\theta},\frac{1}{2}\right) \right\rbrace $. The general solution $(x,m)$ of the system satisfies the following equation
	\begin{equation}\nonumber
		\begin{split}
			\frac{dm}{dx}=\frac{\epsilon m(1-m)(\theta x-(1-x))}{x(1-x)(\phi x+\psi(k))}.
		\end{split}
	\end{equation}
	Hence, $V(x,m)=0$.
	Let $\omega=\left( x',\frac{1}{2}\right) $ be any point on the line $ m = \frac{1}{2} $, which represents one of the $ x $-nullclines (i.e., $\Phi(x',\frac{1}{2})=0$), where $x'\in(0,\frac{1}{1+\theta})$. The solution through $\omega$ spirals around the interior equilibrium point and satisfies the equation
	\begin{equation}\label{G(x)}
		G(x)-\ln(m(1-m))+\ln(\frac{1}{4})-G(x')=0,
	\end{equation}
	which only intersects the line $m=\frac{1}{2}$ at two points, including $\omega$. Therefore, every solution of the system \eqref{rep_eq} is a closed orbit in $D$.
	
	Next, we turn to a special dynamic of the system, the existence of a heteroclinic cycle
	$a_{1}=(0,0)\to a_{2}=(1,0)\to a_{3}=(1,1)\to a_{4}=(0,1)\to a_{1}$; that is, the four boundary saddle points are arranged in a cyclic pattern.
	Actually, each of these four corner equilibria has one positive and one negative transversal eigenvalue.
	On the line $a_{1}a_{2}$, the direction of the vector field is from $a_{1}$ to $a_{2}$, resulting in trajectories starting $ (x,0) $ ($x\in(0,1)$) converging to $a_{2}$. Similarly, there are trajectories from $a_{2}$ to $a_{3}$, $a_{3}$ to $a_{4}$, and $a_{4}$ to $a_{1}$, collectively forming the heteroclinic cycle.

	\subsection{Proof of Theorem \ref{thm2}}\label{A.3}
	
	The condition $T-R<{\rm min}\left\lbrace\frac{P-S}{1-k},(1-k)(P-S) \right\rbrace $ consists of three parts: (1) $ T-R<0 $, $P-S<0$; (2) $P-S<0$, $ T-R>0 $, $ \frac{T-R}{P-S}>\frac{1}{1-k} $; (3) $P-S>0$, $ T-R<0 $, $\frac{T-R}{P-S}<1-k$.
	Additionally, there are four corner equilibrium points: $ (0, 0) $, $ (1, 0) $, $ (1, 1) $,  $ (0, 1) $; and one interior equilibrium point  $\left(\frac{1}{1+\theta},\frac{1}{2}\right)$.
	
	It is evident that $\frac{1}{\epsilon}\left(\delta_{TR}+\frac{\delta_{PS}+\delta_{TR}}{k-2} \right)<0$ and $\frac{1}{\epsilon}\left(\delta_{TR}+\frac{\delta_{PS}+\delta_{TR}}{k-2} \right)<0$. Consequently, each Jacobian of the corner equilibrium points $ (0, 1) $ and $ (1, 0) $ contains two positive eigenvalues. Similarly, each Jacobian of corner equilibrium points $ (0, 0) $ and $ (1, 1) $ has two negative eigenvalues. Therefore, $ (0, 1) $ and $ (1, 0) $ are unstable nodes, while $ (0, 0) $ and $ (1, 1) $ are stable nodes.
	For the interior equilibrium point $\left(\frac{1}{1+\theta},\frac{1}{2}\right)$, the eigenvalues of its Jacobian are
	$$
	\lambda_{1}= \frac{\sqrt{(2k-4)\theta}}{\sqrt{\epsilon}(k-2)(1+\theta)}\sqrt{(\theta k+1-\theta)(\delta_{TR}-\delta_{PS})-(\theta k+k)\delta_{TR}}\ >0
	$$
	and
	$$
	\lambda_{2}=- \frac{\sqrt{(2k-4)\theta}}{\sqrt{\epsilon}(k-2)(1+\theta)}\sqrt{(\theta k+1-\theta)(\delta_{TR}-\delta_{PS})-(\theta k+k)\delta_{TR}}\ <0.
	$$
	Thus $\left(\frac{1}{1+\theta},\frac{1}{2}\right)$ is the saddle point.
	
	\subsection{Proof of Theorem \ref{thm3}}\label{A.4}
	
	For the condition $\frac{\theta k-\theta+1}{1-\theta-k}(P-S)<T-R< {\rm max}\left\lbrace\frac{P-S}{1-k},(1-k)(P-S) \right\rbrace $, there are two cases: (1)  $ T-R>0 $, $ P-S<0 $, $ 1-k<\frac{T-R}{P-S}<\frac{\theta k-\theta+1}{1-\theta-k}$, and (2) $ T-R<0 $,  $P-S>0$, $ \frac{\theta k-\theta+1}{1-\theta-k}<\frac{T-R}{P-S}<\frac{1}{1-k}$.
	Each case leads to seven equilibrium points in the system: the four corner points  $(0,0)$, $(1,0)$, $(1,1)$, $(0,1)$; two boundary points: $\left( \frac{k-1}{k-2}+\frac{k(R-T)}{(k-2)(T-R-P+S)},1\right)$, $\left( \frac{k-1}{k-2}+\frac{k(R-T)}{(k-2)(T-R-P+S)},0\right)$; and  one interior point:
	$\left(\frac{1}{1+\theta},\frac{1}{2}\right)$.
	
	For the interior point $\left(\frac{1}{1+\theta},\frac{1}{2}\right)$, the eigenvalues of its Jacobian are computed as
	$$
	\lambda_{1}= \frac{\sqrt{(2k-4)\theta}}{\sqrt{\epsilon}(k-2)(1+\theta)}\sqrt{(\theta k+1-\theta)(\delta_{TR}-\delta_{PS})-(\theta k+k)\delta_{TR}}
	$$
	and
	$$		
	\lambda_{2}=- \frac{\sqrt{(2k-4)\theta}}{\sqrt{\epsilon}(k-2)(1+\theta)}\sqrt{(\theta k+1-\theta)(\delta_{TR}-\delta_{PS})-(\theta k+k)\delta_{TR}}\ ,
	$$
	both being purely imaginary. By following a similar proof process as in \ref{A.2}, we derive that it is a center.
	
	Furthermore, in the case of $1-k<\frac{T-R}{P-S}<\frac{\theta k-\theta+1}{1-\theta-k}$ with $T-R>0$ and $P-S<0$, we have
	$$
	\frac{1}{\epsilon}\left(\delta_{PS}+\frac{\delta_{PS}+\delta_{TR}}{k-2} \right)<0,\quad \frac{((k+1)\delta_{PS}+\delta_{TR})(\delta_{PS}+(k-1)\delta_{TR})}{\epsilon(k-2)^{2}(\delta_{TR}-\delta_{PS})}<0,
	$$
	$$
	\frac{1}{\epsilon}\left(\delta_{TR}+\frac{\delta_{PS}+\delta_{TR}}{k-2} \right)>0,\quad {\rm and} \quad 1-(\theta+1)\left( \frac{\delta_{TR}+(k-1)\delta_{PS}}{(2-k)(\delta_{TR}-\delta_{PS})}\right)>0.
	$$
	Therefore, $(0,1)$ is an unstable node and $(0,0)$ is a stable node. $(1,0)$, $(1,1)$, $\left( \frac{k-1}{k-2}+\frac{k(R-T)}{(k-2)(T-R-P+S)},1\right)$,  and $\left( \frac{k-1}{k-2}+\frac{k(R-T)}{(k-2)(T-R-P+S)},0\right)$ are unstable saddles.
	
	\vskip 0.3cm
	
	Likewise, in the case of $\frac{\theta k-\theta+1}{1-\theta-k}<\frac{T-R}{P-S}<\frac{1}{1-k}$ with $T-R<0$ and $P-S>0$, we have
	$$
	\frac{1}{\epsilon}\left(\delta_{PS}+\frac{\delta_{PS}+\delta_{TR}}{k-2} \right)>0,\quad \frac{((k+1)\delta_{PS}+\delta_{TR})(\delta_{PS}+(k-1)\delta_{TR})}{\epsilon(k-2)^{2}(\delta_{TR}-\delta_{PS})}>0,
	$$
	$$
	\frac{1}{\epsilon}\left(\delta_{TR}+\frac{\delta_{PS}+\delta_{TR}}{k-2} \right)>0,\quad {\rm and} \quad
	1-(\theta+1)\left( \frac{\delta_{TR}+(k-1)\delta_{PS}}{(2-k)(\delta_{TR}-\delta_{PS})}\right)<0.
	$$
	Hence, $(1,1)$ is a stable node and $(1,0)$ is an unstable node. Additionally,  $ (0, 0) $, $ (0, 1) $, $\left( \frac{k-1}{k-2}+\frac{k(R-T)}{(k-2)(T-R-P+S)},0\right)$, and $\left( \frac{k-1}{k-2}+\frac{k(R-T)}{(k-2)(T-R-P+S)},1\right)$ are unstable saddles.

	Furthermore, similar to the proof of Theorem \ref{thm1},  in the situation where $T-R<{\rm min}\left\lbrace\frac{P-S}{1-k},(1-k)(P-S) \right\rbrace, $
	there exists a heteroclinic cycle as follows:
	(1) For the case of $P-S<0$ and $T-R>0$: $(1,0)$ $\to$ $(1,1)$ $\to$ $\left( \frac{k-1}{k-2}+\frac{k(R-T)}{(k-2)(T-R-P+S)},1\right)$ $\to$ $\left( \frac{k-1}{k-2}+\frac{k(R-T)}{(k-2)(T-R-P+S)},0\right)$ $\to$ $(1,0)$;
	(2) For the case of $P-S>0$ and $T-R<0$: $ (0,0)$ $ \to$ $\left( \frac{k-1}{k-2}+\frac{k(R-T)}{(k-2)(T-R-P+S)},0\right)$ $\to$ $\left( \frac{k-1}{k-2}+\frac{k(R-T)}{(k-2)(T-R-P+S)},1\right)$ $\to$ $(0,1)$ $\to$ $(0,0)$.

	\subsection{Proof of Theorem \ref{thm4}}\label{A.5}
	
	When  ${\rm min}\left\lbrace\frac{P-S}{1-k},(1-k)(P-S)\right\rbrace<T-R<\frac{\theta k-\theta+1}{1-\theta-k}(P-S)$, there are two cases: (1) $T-R>0$, $P-S<0$, $ \frac{\theta k-\theta+1}{1-\theta-k}<\frac{T-R}{P-S}<\frac{1}{1-k}$, and (2) $T-R<0$,  $P-S>0$, $ 1-k<\frac{T-R}{P-S}<\frac{\theta k-\theta+1}{1-\theta-k}$.
	Additionally, the system has seven equilibrium points: $\left(\frac{1}{1+\theta},\frac{1}{2}\right)$, $(0,0)$, $(1,0)$, $(1,1)$, $(0,1)$, $\left( \frac{k-1}{k-2}+\frac{k(R-T)}{(k-2)(T-R-P+S)},1\right)$, and $\left( \frac{k-1}{k-2}+\frac{k(R-T)}{(k-2)(T-R-P+S)},0\right)$.
	
	For the interior point $\left(\frac{1}{1+\theta},\frac{1}{2}\right)$, the eigenvalues of its Jacobian are given by
	$$
	\lambda_{1}= \frac{\sqrt{(2k-4)\theta}}{\sqrt{\epsilon}(k-2)(1+\theta)}\sqrt{(\theta k+1-\theta)(\delta_{TR}-\delta_{PS})-(\theta k+k)\delta_{TR}}\ >0
	$$
	and
	$$
	\lambda_{2}=- \frac{\sqrt{(2k-4)\theta}}{\sqrt{\epsilon}(k-2)(1+\theta)}\sqrt{(\theta k+1-\theta)(\delta_{TR}-\delta_{PS})-(\theta k+k)\delta_{TR}}\ <0,
	$$
	indicating that $\left(\frac{1}{1+\theta},\frac{1}{2}\right)$ is a saddle point.
	
	Furthermore, when  $T-R>0$, $P-S<0$, and $\frac{\theta k-\theta+1}{1-\theta-k}<\frac{T-R}{P-S}<\frac{1}{1-k}$,
	we have
	$$
	\frac{1}{\epsilon}\left(\delta_{PS}+\frac{\delta_{PS}+\delta_{TR}}{k-2} \right)<0,\quad 1-(\theta+1)\left( \frac{\delta_{TR}+(k-1)\delta_{PS}}{(2-k)(\delta_{TR}-\delta_{PS})}\right)<0,
	$$
	$$ \frac{1}{\epsilon}\left(\delta_{TR}+\frac{\delta_{PS}+\delta_{TR}}{k-2} \right)>0,\quad {\rm and} \quad \frac{((k+1)\delta_{PS}+\delta_{TR})(\delta_{PS}+(k-1)\delta_{TR})}{\epsilon(k-2)^{2}(\delta_{TR}-\delta_{PS})}<0.
	$$
	Therefore, the equilibrium points $(1, 0) $ and $ (1, 1) $ are saddle points, the equilibrium points $(0,0)$ and $\left( \frac{k-1}{k-2}+\frac{k(R-T)}{(k-2)(T-R-P+S)},1\right)$ are stable nodes, while  $ (0, 1) $ and $\left( \frac{k-1}{k-2}+\frac{k(R-T)}{(k-2)(T-R-P+S)},0\right)$ are unstable nodes.
	
	\vskip 0.3cm
	
	Similarly, when $T-R<0$, $P-S>0$, and $ 1-k<\frac{T-R}{P-S}<\frac{\theta k-\theta+1}{1-\theta-k} $, the equilibrium points $(0,0)$ and $(0,1)$ are saddle points;
	$(1,0)$ and $\left( \frac{k-1}{k-2}+\frac{k(R-T)}{(k-2)(T-R-P+S)},1\right)$ are unstable nodes; $ (1, 0) $ and $\left( \frac{k-1}{k-2}+\frac{k(R-T)}{(k-2)(T-R-P+S)},0\right)$ are stable nodes.

\end{document}